\documentclass[journal]{IEEEtran}
\usepackage{amsmath,amsfonts}
\usepackage{algorithmic}
\usepackage{algorithm}
\usepackage{array}
\usepackage[caption=false,font=normalsize,labelfont=sf,textfont=sf]{subfig}
\usepackage{textcomp}
\usepackage{stfloats}
\usepackage{url}
\usepackage{verbatim}
\usepackage{graphicx}
\usepackage{cite}
\usepackage{orcidlink}
\usepackage{multirow}
\usepackage{enumitem}

\begin{document}

\newcommand{\figureref}[1]{Figure \ref{#1}}
\newcommand{\tableref}[1]{Table \ref{#1}}
\newcommand{\equationref}[1]{Equation \ref{#1}}
\newcommand{\sectionref}[1]{Section \ref{#1}}

\newcommand{\textcite}[1]{\cite{#1}}
\newcommand{\parencite}[1]{\cite{#1}}

\newcommand{\change}[1]{\textcolor{red}{#1}}
\newcommand{\missing}[1]{\textcolor{green}{#1}}
\newcommand{\new}[1]{\textcolor{blue}{#1}}


\title{Mitigation of multi-path propagation artefacts in acoustic targets with adaptive cepstral filtering}

\author{
Lucas C. F. Domingos\orcidlink{0000-0001-8205-1562},
Paulo E. Santos\orcidlink{0000-0001-8484-0354},
Karl Sammut\orcidlink{0000-0002-7654-9978}~\IEEEmembership{Senior Member,~IEEE,},
and Russell S. A. Brinkworth\orcidlink{0000-0003-0270-3538}
\thanks{Manuscript received XXX XX, 2025; revised XXX XX, 2025. This work was supported by the National Industry PhD Program, in a partnership between Flinders University and Priori Analytica. \emph{(Corresponding author: Lucas C.F. Domingos.)}}%
\thanks{Lucas C. F. Domingos, Russell S. A. Brinkworth, and Karl Sammut are with the Centre for Defence Engineering Research and Training, College of Science and Engineering, Flinders University, Tonsley, Australia (e-mail:lucas.domingos@flinders.edu.au; russell.brinkworth@flinders.edu.au; karl.sammut@flinders.edu.au).}
\thanks{Paulo E. Santos is with the Department of Research And Development, PrioriAnalytica, Adelaide, Australia (e-mail:paulo.santos@priorianalytica.com).}
}

\markboth
{\MakeUppercase{Domingos} \MakeLowercase{\textit{et al.}}: \MakeUppercase{Mitigation of multi-path propagation artefacts in acoustic targets with cepstral adaptive filtering}}
{IEEE TRANSACTIONS ON GEOSCIENCE AND REMOTE SENSING}

\maketitle

\begin{abstract}
Passive acoustic sensing is a cost-effective solution for monitoring moving targets such as vessels and aircraft, but its performance is hindered by complex propagation effects like multi-path reflections and motion-induced artefacts. 
Existing filtering techniques do not properly incorporate the characteristics of the environment or account for variability in medium properties, limiting their effectiveness in separating source and reflection components. 
This paper proposes a method for separating target signals from their reflections in a spectrogram. Temporal filtering is applied to cepstral coefficients using an adaptive band-stop filter, which dynamically adjusts its bandwidth based on the relative intensity of the quefrency components.
The method improved the signal-to-noise ratio (SNR) and log-spectral distance (LSD) across velocities ranging from 10 to 100 metres per second in aircraft noise with simulated motion. It also enhanced the performance of ship-type classification in underwater tasks by 2.28 and 2.62 Matthews Correlation Coefficient percentage points for the DeepShip and VTUAD v2 datasets, respectively. 
These results demonstrate the potential of the proposed pipeline to improve acoustic target classification and time-delay estimation in multi-path environments, with future work aimed at amplitude preservation and multi-sensor applications.

\end{abstract}


\section{Introduction}
\label{sec:introduction}

\IEEEPARstart{T}{he} classification and localisation of moving acoustic sources are critical for applications in surveillance, environmental monitoring, ecology, and collision mitigation. Vessels, aircraft, and wildlife animals are common targets of these applications \parencite{fang2022acoustic, dong2025cafvit}. For those scenarios, passive acoustic sensing offers a cost-effective and non-intrusive solution, especially when using single-sensor setups \parencite{ahmad2024deep}. However, the sound propagation of moving targets is complex and poses a challenge to such systems. 

The target signature varies with the movement configuration and surrounding environment, depending on factors such as the distance and grazing angle of approach \parencite{sanchez-perez2013aircraft}, latitude and longitude effects \parencite{leroy2008new}, and the geometry of the environment \parencite{urick1979sound}. These factors cause phenomena that further contribute to undesirable distortions, such as multi-path reflections, non-linear attenuation, and frequency shifts, the latter often resulting from the Doppler effect, characterised by the compression and stretching of the wavefront \parencite{rossing2014springer,audoly2017measurement,li2024mitigation}. When combined with detection or classification tasks, these factors pose a significant challenge to acoustic sensing solutions, highlighting the need for adaptive processing that can adjust its characteristics based on the received stimulus.

A central acoustic phenomenon leveraged in passive acoustic sensing applications is the multi-path reflection caused by interference between direct and surface-reflected sound paths. When a sound-emitting source is in motion, the Lloyd’s Mirror Effect (LME) \parencite{jensen2011computational, rodriguez2023fundamentals} is observed through constructive and destructive interference patterns produced by the propagated sound waves. Despite the apparent degradation of the signal, these distinct patterns contain valuable information about the physical characteristics of the target and its motion properties, encoding factors such as velocity, distance, and medium properties. Various applications arise from the study of these patterns, including environmental modelling and acoustic propagation studies \parencite{audoly2017measurement}, vessel tracking and localisation \parencite{ferguson2017convolutional, gaggero2024possibility}, marine mammal depth estimation \parencite{bouffaut2021source, pereira2020use}, and aircraft flight parameter estimation \parencite{lo2002aircraft}.

Since LME patterns are readily identifiable in spectrograms, the depth of the acoustic source encoded in LME can be estimated when prior knowledge of receiver positioning and expected frequency components are available. In marine ecology, LME patterns are used as references for estimating the vocalisation depth of whales \parencite{pereira2020use, bouffaut2021source}. In surveillance applications, the same principles apply to the underwater acoustic localisation of moving targets \parencite{zhao2024singlehydrophonebased, meng2016motion} and the estimation of aircraft flight parameters \parencite{lo2002aircraft}. However, the presence of this phenomenon can also be undesirable in certain contexts. In the underwater domain, for instance, multi-path reflections are known to contribute to ship collisions with marine mammals, as their perception of sound is disrupted by such interactions \parencite{gerstein2005acoustics}.
Similarly, these effects increase complexity in classification and recognition tasks within surveillance and defence applications, as the resulting patterns can cause spectral ambiguity, hindering the identification of the true spectral components \parencite{domingos2022survey}.

Cepstral analysis has proven to be a robust technique for extracting meaningful features from multi-path reflection patterns, as this representation separates the signal and its reflections in an additive manner \parencite{oppenheim2004frequency}. The cepstrum is obtained by applying the inverse Fourier transform to the logarithm of the signal’s spectrum. The terminology used to describe cepstral components is adapted from spectral analysis: the counterparts of frequency components are referred to as quefrency, while the counterparts of harmonic components are referred to as rahmonics. Analysis of this representation reveals periodic structures caused by multi-path delays \parencite{bogert1963quefrency}. Applications involving cepstral analysis have demonstrated superior performance compared to other techniques, such as correlation analysis, in multi-path environments \parencite{gao2008time}, which is characteristic of scenarios where LME is present. Additionally, cepstral features are commonly used as input in machine learning applications, enabling accurate range prediction and classification \parencite{ferguson2017convolutional, abdul2022mel}.

Spectral filtering can be applied to separate multi-path reflection components from acoustic signals \parencite{yakubovskiy2015feature, zhao2024singlehydrophonebased}; however, filtering in the cepstral domain presents a strong, and often more robust, alternative. A solution was proposed for time-of-arrival estimation of ship sounds in shallow waters, which involved separating the upper and lower bounds of the cepstrogram, corresponding to low and high quefrencies, from the intermediate quefrency bands. Using the mid-quefrency band, the spectrogram was reconstructed, enabling the effective removal of ship-specific components that were undesirable for the task \parencite{ferguson2017convolutional}. A similar strategy was applied for the detection of unmanned aerial vehicles, retaining only the harmonic components associated with target-specific noise \parencite{fang2022acoustic}. Both approaches indicate that harmonic components from the target are separable from reflected components through cepstral analysis, although they rely on techniques that remove entire quefrency bands without prior identification of the components.

Improvements to cepstral separation were achieved by averaging the power cepstrum over a sufficiently long time window, providing an estimate of the non-rahmonic components (i.e., those not resulting from multi-path interactions), which could then be subtracted from the original cepstrum. This technique, which was named cepstrum subtraction \parencite{ferguson2019improved}, has proven effective in enhancing the extraction of rahmonic components for time-delay estimation. Similarly, the removal of cepstral local maxima was applied to reduce multi-path artefacts in hydroacoustic channels \parencite{czapiewska2020reduction}, improving data transmission quality by up to five times. Both approaches enhance the separation of multi-path components by estimating the behaviour of the rahmonics; however, when applied across the entire quefrency range, they incorporate bands into the filtering process that are not influenced by these artefacts. Moreover, they do not account for the physical constraints of reflection patterns, which could further improve the robustness of such applications.

Despite existing cepstral filtering techniques aimed at separating reflected components from acoustic data, further improvements can be achieved by incorporating physical limitations of the environment into the filtering process. Additionally, current methods often do not account for variability in the target signature arising from movement configuration and environment, thereby leaving scope for adaptive filtering of cepstral representations.

In this paper, a temporal filtering approach for spectrograms based on cepstral coefficients is proposed, incorporating an adaptive band-stop filter in the cepstral domain. This filter dynamically adjusts its strength based on the relative intensity of individual quefrency components, enabling more effective suppression of undesired reflections while preserving target-specific harmonic features. The main contributions of this paper are as follows:

\begin{enumerate}
\item An innovative temporal filtering method for cepstral representations, including an adaptive band-stop filter designed to reduce the effects of multi-path propagation in acoustic channels by leveraging the relative intensity of quefrency components.
\item A robust evaluation of the proposed filtering method, encompassing both the removal of multi-path reflections in simulated data and their attenuation in a practical classification task.
\end{enumerate}

To support the proposed method, a formal introduction to the Lloyd’s Mirror Effect (LME) in acoustics is provided in \sectionref{sec:cepstral}. The method for attenuating motion artefacts, based on temporal filtering of the cepstrogram and the proposed adaptive band-stop filtering technique, is introduced in \sectionref{sec:methods}. A detailed description of the experiments used to evaluate the method is presented in \sectionref{sec:experimental}, which includes two scenarios: attenuation of movement artefacts in real data with simulated motion (\sectionref{subsec:simulated}), and classification of underwater acoustic signals using real-world recordings affected by LME (\sectionref{subsec:classification}). The corresponding results and analysis are discussed in \sectionref{sec:results}. Finally, future directions and concluding remarks are presented in \sectionref{sec:final_remarks}.

\section{Lloyd's Mirror Effect in Acoustic Recordings}
\label{sec:cepstral}
\begin{figure}[htb]
    \centering
    \resizebox{0.95\columnwidth}{!}{\includegraphics{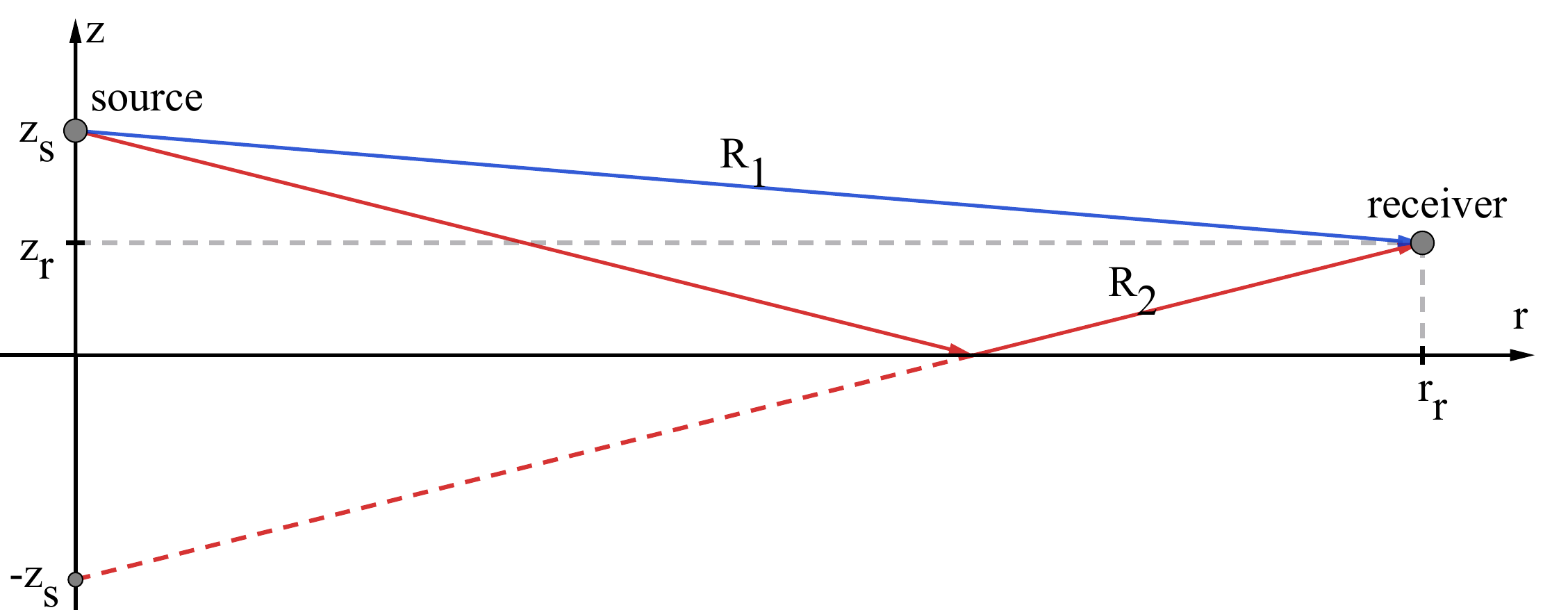}}   
    \caption[Reflection of an acoustic source on a bounded medium.]{Reflection of an acoustic source on a bounded medium. $z_{s}$ and $z_{r}$ represent the perpendicular distances from the source and the receiver to the boundary, respectively; $r_{r}$ denotes the horizontal distance between the source and the receiver.}
    \label{fig:lme_reflection}
\end{figure}

An acoustic transducer acting as a receiver measures the time-varying pressure field at its proximity. Given a spherical wave propagating from a source, the pressure field $p$ as a function of the radial distance $r$ can be described as

\begin{equation}
    p = A \frac{e^{jkR}}{R},
    \label{eq:spherical_spreading}
\end{equation}

\noindent where $A$ is the amplitude of the sound source, $R$ denotes the distance between the source and receiver, and the wave number is given by $k = \frac{2\pi f}{c}$ \parencite{rossing2014springer}. In an homogenetic medium bounded with a perfectly reflective surface, the sound waves emitted by the source reflect at the boundaries, and the pressure field measured at the receiver is the combination of the direct and reflected components, as illustrated in \figureref{fig:lme_reflection}. This pressure field can be described as

\begin{equation}
    p = \frac{e^{ikR_1}}{R_1} - \frac{e^{ikR_2}}{R_2},
    \label{eq:lme_equation}
\end{equation}

\noindent where 

\begin{align*}
    R_1 = \sqrt{r_{r}^2 + (z_s-z_r)^2} && \text{and} && R_2 = \sqrt{r_{r}^2 + (z_s+z_r)^2},
\end{align*}

\noindent with $r_{r}$ representing the horizontal distance between the source and receiver; $z_s$ and $z_r$ representing, respectively, the vertical distance from the medium to the source and receiver.

Variations in the distance $r_{r}$ produce a complex interference pattern, known as Lloyd's Mirror Effect (LME) \parencite{jensen2011computational,rodriguez2023fundamentals}, resulting from the constructive and destructive interference caused by multi-path reflections of the source wave. The spectrum of a broadband source recorded under these conditions exhibits this complex pattern across all wave numbers, with its characteristics dependent on the properties of the medium. It is important to note that variations in the pattern occur under non-ideal conditions, such as irregular surfaces, multiple boundaries, or stratified media, requiring adaptations to the original formulation in extreme scenarios. In applications where the source information is of interest rather than the medium properties, the LME may represent an undesirable component in the received signal. 

This work focuses on the mitigation of the LME in acoustic recordings. Given a scenario with fixed receiver, the interference pattern recorded will be a function of $z_{s}$ and $r_{r}$, as indicated in \equationref{eq:lme_equation}. In a target detection application, the removal or attenuation of the LME can be beneficial. For aircraft detection, for instance, the removal of the LME will decrease the influence of the height ($z_{s}$) and distance to receiver ($r_{r}$), potentially contributing to a better identification of the source. However, in areas such as surface ships classification using underwater acoustic signals, the complete removal of LME will also remove the influence of the source depth ($z_{s}$), which may contain valuable information about the vessel size and tonnage, thus, the mitigation is more appropriate to decrease the influence of the distance to receiver ($r_{r}$). For both applications, the analysis of the cepstral components serves as a valuable tool for addressing echoes and delays produced by movement in acoustic recordings. The following section introduces a method for filtering the cepstrogram to separate reflection components from the signal spectrogram.

\section{Methods}
\label{sec:methods}
\subsection{Time-filtering of the Cepstrogram}
\begin{figure}[tb]
    \centering
    \resizebox{\columnwidth}{!}{\includegraphics{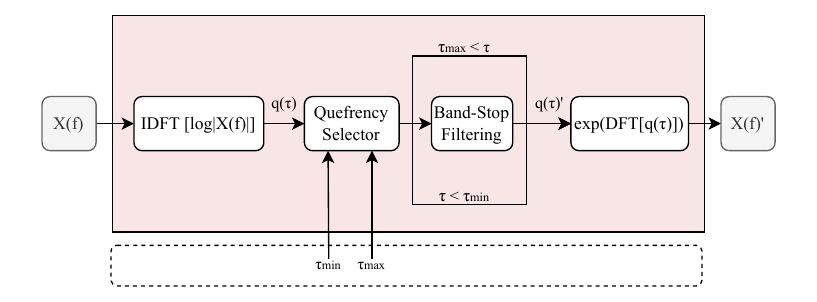}}   
    \caption[A block diagram representing the steps of the cepstrogram time-filtering method proposed.]{A block diagram representing the steps of the proposed cepstrogram time-filtering method. The band stop filter operates in the time-quefrency domain, processing each quefrency individually along the time axis.}
    \label{fig:cepstrum_filt_bd}
\end{figure}

Given an acoustic recording containing a broadband signal affected by LME, the influence of reflections can be mitigated by analysing its cepstral representation. In this work, a method is proposed for filtering the cepstral representation to attenuate these effects. The complete pipeline of the proposed filtering method consists of four stages, as illustrated in \figureref{fig:cepstrum_filt_bd}. The first stage involves extracting the cepstral components from the spectral representation. Next, the quefrencies affected by reflections are identified and selected. These selected components are then filtered along the time axis. Finally, the filtered cepstral representation is transformed back into the spectral domain.

To obtain the cepstral representation, the spectrogram is initially obtained. The Short-Time Fourier Transform (STFT) \parencite{rossing2014springer} is commonly used to generate the spectrogram; however, approaches such as Mel frequency spectrograms \parencite{stevens1937scale}, which produce log-scaled frequencies that mimic human auditory perception, and the Constant Q Transform (CQT) \parencite{brown1991calculation}, which enhances frequency resolution using geometrically spaced filters, are preferred in domain-specific applications. 

Once the time-frequency representation of the signal has been obtained, the cepstral representation needs to be extracted. Following the convention \parencite{rossing2014springer}, the cepstrum is obtained by performing the inverse Fourier transform of the log-spectrum magnitude, as described by
\begin{equation}
    q(\tau) = IDFT \left[ \log|X(f)| \right],
    \label{eq:cepstrum_calculation}
\end{equation}

\noindent where $\tau$ denotes the quefrency, $|X(f)|$ is the spectrum magnitude, and $IDFT$ denotes the Inverse Discrete Fourier Transform \parencite{rossing2014springer}. Analogously, the cepstrogram is formed by concatenating the cepstra obtained for all spectra in the initial spectrogram. It is worth noting that this formulation corresponds to the real cepstrum, which is derived from the log-magnitude spectrum and therefore does not retain phase information. While phase information can encode fine temporal structure, magnitude-based cepstral representations are known to effectively capture spectral periodicities associated with harmonics and reflections \parencite{randall2021cepstrum}.

Considering the geometry of a source with constant $z_{s}$ and $r_{r}$ varying over time, the quefrency $\tau$, which is a time variable representing the reception delay, reaches its maximum value at the minimum distance $r_{r}$. Thus, quefrencies greater than the maximum value $\tau_{max}$ are not produced by LME and are therefore excluded from this analysis. Conversely, as the quefrency approaches zero (the lower part of the cepstrum), the information it contains is highly source-dependent, since the time delay is close to zero. Therefore, lower quefrencies below a minimum $\tau_{min}$ are also ignored during the filtering process. The cepstral band affected by the LME, corresponding to the central quefrencies of the cepstrum, must be processed to mitigate the effect. 

The cepstrogram of a moving target will contain quefrency peaks at periodicity determined by the target's speed. Therefore, assuming a known velocity range, the cepstrum of the pattern described by \equationref{eq:lme_equation} contains a magnitude peak at the quefrency $\tau$ of limited duration, which can be filtered out by applying a discrete band-stop filter along the time axis of the cepstrogram.  In practice, the filtering process separates the direct path signal from its reflected components, which can then be removed. After the cepstrogram filtering, the spectrogram is recovered by applying the inverse operations from \equationref{eq:cepstrum_calculation}, in the form 
\begin{equation}
  |X(f)| = e^{DFT \left[ q(\tau)\right]},
    \label{eq:icepstrum_calculation}
\end{equation}

where $DFT$ denotes the Discrete Fourier Transform \parencite{rossing2014springer}. 

A fundamental step in the cepstral filtering proposed in this paper is the application of a band-stop filter in the quefrency components. Improvements in this stage can greatly contribute to the  performance of the filtering, enhancing its capability of distinguishing between reflection components and the signal features. An adaptive bandwidth selection based on signal features is then proposed for this pipeline.

\subsection{Adaptive Band-Stop Filtering}

A band-stop filter is composed from a combination of a low-pass and a high-pass filter. A discrete single-pole infinite impulse response (IIR) low-pass filter ($LPF$) can be described as
\begin{equation}
    LPF(x_{t}, y_{t-1}) = \frac{2\pi f_c}{f_r} x_t + \left( 1 - \frac{2\pi f_c}{f_r} \right) y_{t-1},
\label{eq:fixed_lpf}
\end{equation}

\noindent where $f_c$ denotes the 3dB corner frequency of the filter, $f_r$ is the sample rate, $x_{t}$ is the current discrete point, and $y_{t-1}$ is the output from the previous step. In terms of the low-pass filter, a high-pass filter can be described as

\begin{equation}
    HPF(x_{t}, y_{t-1}) = x_{t} - LPF(x_{t}, y_{t-1}).
\end{equation}

Accordingly, a band-stop filter (BSF) dependent on the low-pass corner frequency $f_{c1}$, the high-pass corner frequency $f_{c2}$, and the sample rate $f_r$, can be described as 

\begin{equation}
    BSF(x_{t}, y_{t-1}) = x_{t} + LPF_{1}(x_{t}, y_{t-1}) - LPF_{2}(x_{t}, y_{t-1}).
    \label{eq:fixed_bandstop}
\end{equation}

The BSF, which comprises a pair of IIR low-pass filters, exhibits a fixed frequency response. Consequently, its behaviour is independent of the input signal and insensitive to variations in input features.

Despite the simplicity of the frequency filter, real-world recordings often deviate from ideal assumptions, exhibiting variations in target speed, irregularities in the medium boundary, and inhomogeneities within the medium \parencite{audoly2017measurement}. In such cases, a fixed corner frequency can be limiting, and adaptive bandwidth selection based on signal features is preferred. As the LME pattern appears in the cepstrogram as quefrency peaks, the presence of reflected components is directly proportional to the cepstral magnitude. Therefore, the bandwidth of the band-stop filter should adapt to amplitude changes, broadening as the amplitude increases and narrowing as it decreases. Thus, when strong multi-path components are present, the filter removes their influence across a wider frequency band.

\begin{figure}[ht]
    \centering
    \resizebox{\columnwidth}{!}{\includegraphics{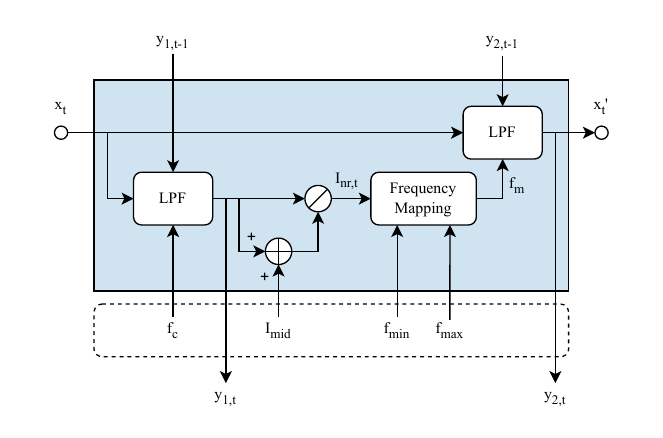}}   
    \caption[A block diagram representing the low-pass adaptive filtering process.]{A block diagram representing the low-pass adaptive filtering process. The variables in the dashed box indicate the parameters of the filter.}
    \label{fig:adap_filt_bd}
\end{figure}

To implement a filter with adaptive corner frequencies, the variable low-pass filter, initial stage of the motion processing model \parencite{brinkworth2009robust}, was employed. This filter applies a nonlinearity to the signal and uses the output to select a suitable corner frequency based on a reference range. The process, illustrated in \figureref{fig:adap_filt_bd}, can be divided into four steps. 

The first step is the application of an IIR low-pass filter to the signal, following \equationref{eq:fixed_lpf}, to increase the SNR. By removing high-frequency components, noise is reduced in the representation. Then, to estimate the adaptation state, the result passes through a tone-mapping operation following the Naka-Rushton transform \parencite{naka1966spotentials}, expressed as

\begin{equation}
    I_{nr,t} = \frac{y_{1,t}}{y_{1,t}+I_{mid}},
\end{equation}

\noindent where $y_{1,t}$ is the output of the low-pass filter, and $I_{mid}$ is an empirical reference value. Using $I_{mid}$ as a reference, the relative amplitude of the input signal can be estimated, which is proportional to the required filtering. The corner frequency is then selected according to the adaptation state. Considering a band-stop filter with corner frequencies $f_{m_1} < f_{m_2}$, as stronger filtering is needed when the LME is present, the adaptive low-pass corner frequency $f_{m_{1}}$ is defined as

\begin{equation}
    f_{m_{1}} = f_{max_{1}} - (f_{max_{1}}-f_{min_{1}}) \times I_{nr,t},
\end{equation}
 
\noindent where $f_{max_{1}}$ and $f_{min_{1}}$ are the maximum and minimum frequencies describing the lower-frequency range. Similarly, the adaptive high-pass corner frequency $f_{m_{2}}$ is defined as

\begin{equation}
    f_{m_{2}} = (f_{max_{2}}-f_{min_{2}}) \times I_{nr,t} + f_{min_{2}},
\end{equation}

\noindent where $f_{max_{2}}$ and $f_{min_{2}}$ are the maximum and minimum frequencies of the high-frequency range. To ensure the correct definition of the band-stop filter, the condition $f_{max_{1}} < f_{min_{2}}$ must be satisfied. Once the condition holds, the filter band widens for amplitude peaks and narrows for lower values. Accordingly, the filtering strategy adapts to the strength of the reflected components. 

Considering the cepstral-filtering approach, the adaptive band-stop filtering distinguishes between the signal’s cepstral components and its reflections based on amplitude, adjusting the filtering strength accordingly. This approach is particularly beneficial in environments with changing conditions, such as imperfections in the reflection surface or variations in medium properties. In \figureref{fig:cepstrum_filt_bd}, the temporal band-stop filter was replaced by the adaptive-frequency band-stop filter. To evaluate the performance and applicability of the proposed filtering strategy, both simulated and real-world applications are considered in this research.

\section{Experiments}
\label{sec:experimental}
To evaluate the performance of the filtering strategy, two scenarios were selected to provide complementary insights. The first experiment, introduced in \sectionref{subsec:simulated}, focuses on the removal of movement artefacts in a simulated environment, enabling direct comparisons with ground truths, facilitating the filtering performance analysis. The second experiment, described in \sectionref{subsec:classification}, analyses the impact of filtering in a practical application, specifically the attenuation of LME in ship-radiated noise for underwater acoustic classification. The findings from both experiments offer a comprehensive overview of the filtering approach, assessing its effectiveness in both simulated and real-world scenarios.

\subsection{Acoustic Data with Simulated Movement}
\label{subsec:simulated}

In order to evaluate the quality of the filtering pipeline proposed, a set of paired information containing an acoustic signal and its version affected movement artefacts was generated. Spherical spreading behaviour and reflection artefacts were synthetically generated for aircraft machinery sounds, simulating the movement of targets in a bounded media. The machinery sounds were manually selected from the Audio Set dataset \parencite{gemmeke2017audioa}, a database of manually annotated audio events. The Audio Set dataset contains 10-second sound clips drawn from videos, with a total of 527 target classes. The data chosen for this pipeline was recordings of aircraft in idling state, aiming for sounds produced by targets in stationary state to be used as reference. Given the quality variability and the subjectivity of the labels, the samples were manually selected prioritising sounds without interference from other targets (e.g. human speaking, environmental noise, background music). After the selection, 30 audio segments were selected and resampled to 32000 samples per second to ensure the same frequency range for the spectrograms. 

To simulate the reflection and spherical spreading, delays were synthetically added to the reference audio signal. Following the scheme illustrated in \figureref{fig:lme_reflection} and considering a uniform linear motion with source velocity $v_s$, the time of arrival $\sigma_1$ of the sound wave travelling in the direct path can be described as 

\begin{equation}
    \sigma_1 = \frac{\sqrt{(r_0 + v_s \times t)^2 + (z_s-z_r)^2}}{c},
\end{equation}

\noindent where $r_0$ is the initial position of the source, $t$ is the time axis, and $c$ is the speed of sound in the propagation media. An analogous approach was considered for the time of arrival $\sigma_2$ of the reflected wave, resulting in the following equation:

\begin{equation}
    \sigma_2 = \frac{\sqrt{(r_0 + v_s \times t)^2 + (z_s+z_r)^2}}{c}.
\end{equation}

The amplitudes of the direct path propagated and reflected waves were normalised by the distances, following the \equationref{eq:spherical_spreading}. The resulting signal is the direct sum of both components, with the reflected one being multiplied by the reflection coefficient. 

For simulation purposes, the height of the source ($z_s$) and the height of the receiver ($z_r$) were considered as 100 meters and 0.5 meter above the ground, respectively, with speed of sound in the air being adopted as 343 meters per second \parencite{kinsler2000fundamentals}. For each of the aircraft machinery audios present in the dataset, three synthetic versions were produced: one affected by amplitude attenuation only, not considering the time of arrival and, therefore, without delays; one containing only the wave propagated trough the direct path, assuming that there are no reflections; and one with the direct path and reflected components of the audio, simulating the bounded media interaction. For simplicity, the spectrogram of those synthetic versions are denoted in this work as $R(f,t)$, $A(f,t)$ and $B(f,t)$, respectively. 

\begin{figure}[ht]
    \centering
    \resizebox{\columnwidth}{!}{\includegraphics{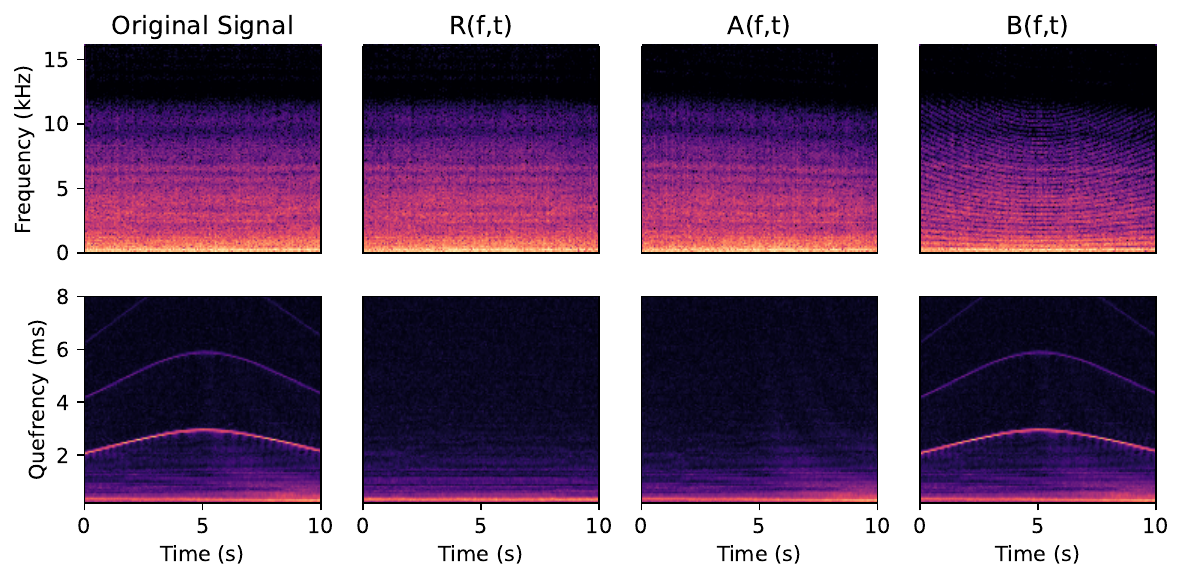}}   
    \caption[Comparison of the spectrograms and cepstrograms of the synthetic dataset components.]{Comparison of the spectrograms (upper row) and cepstrograms (bottom row) of the synthetic dataset components. First column represents the original sample from the Audio Set dataset, $R(f,t)$ is the version with amplitude attenuation due to uniform linear motion, $A(f,t)$ contains the component obtained from the direct path propagation, and $B(f,t)$ contains both the components propagated by the direct and reflected paths.}
    \label{fig:synth_data}
\end{figure}

In order to evaluate the impact of the source velocity ($v_s$) on the filtering strategy, ten evenly spaced velocities were chosen in the range between 10 and 100 meters per second. The final set was then composed of the three synthetic versions for each velocity, considering each of the audios present in the dataset, totalising 900 audios. \figureref{fig:synth_data} illustrates one example of the original audio and the three synthetic versions generated from the pipeline.

To evaluate the performance of the proposed filtering strategy, $A(f,t)$ and $B(f,t)$ were submitted to the cepstral filtering pipeline. The filtered representations were then compared against $R(f,t)$ using two complementary indicators, aiming to a comprehensive performance evaluation of the filtering approach. The first indicator was the Signal-to-Noise Ratio (SNR), which is used to measure the ratio between the reference spectrogram and noise, considering the latter as the difference between the reference spectrogram and the noisy spectrogram. Thus, the SNR is defined as

\begin{equation}
    SNR = 10\times \log\left(\frac{\sum_{f=1}^{F}{\sum_{t=1}^{T}{R(f,t)^2}}}{\sum_{f=1}^{F}{\sum_{t=1}^{T}{[N(f,t) - R(f,t)]^2}}}\right),
    \label{eq:snr}
\end{equation}

\noindent where $R(f,t)$ is the reference spectrogram, $N(f,t)$ is the spectrogram containing both the signal and movement interferences, and $T$ and $F$ denote the total number of time frames and frequency bins, respectively. Higher values of SNR are desirable after a filtering procedure; however, due to its sensitivity to the absolute magnitude values, amplitude variations in the spectrograms can dramatically decrease the SNR.

To offer a more stable measure of quality in the spectrograms, the Log-Spectral Distance (LSD), which is given by

\begin{equation}
    LSD = \frac{1}{F \times T}\sqrt{ \sum_{f=1}^{F}{\sum_{t=1}^{T}{\{\log[N(f,t)] - \log[R(f,t)]\}^2}} },
    \label{eq:lsd}
\end{equation}

\noindent was used as complementary metric. LSD is designed to measure the spectral distortion of the signal, meaning that lower values are associated with improved quality. By considering the logarithmic compression of the signal, LSD provides a perceptually meaningful evaluation of the distortion between two spectral distributions. It is also less sensitive to amplitude variations and take spectral shape variations in consideration \parencite{gray1980distortion}. Both LSD and SNR were used as indicators of the filtering quality in terms of noise attenuation and quality improvement.

\subsection{Underwater Acoustic Signal Classification}
\label{subsec:classification}

Although the performance evaluation applied to acoustic signals affected by simulated movement artefacts provides an indication of the filtering performance, the analysis on real-world data can provide a more practical evaluation of the pipeline. By analysing an application involving signals affected by movement artefacts in real-world conditions, the conclusions are more translatable to practical applications. To address this issue, the task of automatic classification of ship-type using underwater acoustic signals was considered. 

\begin{figure}[ht]
    \centering
    \resizebox{\columnwidth}{!}{\includegraphics{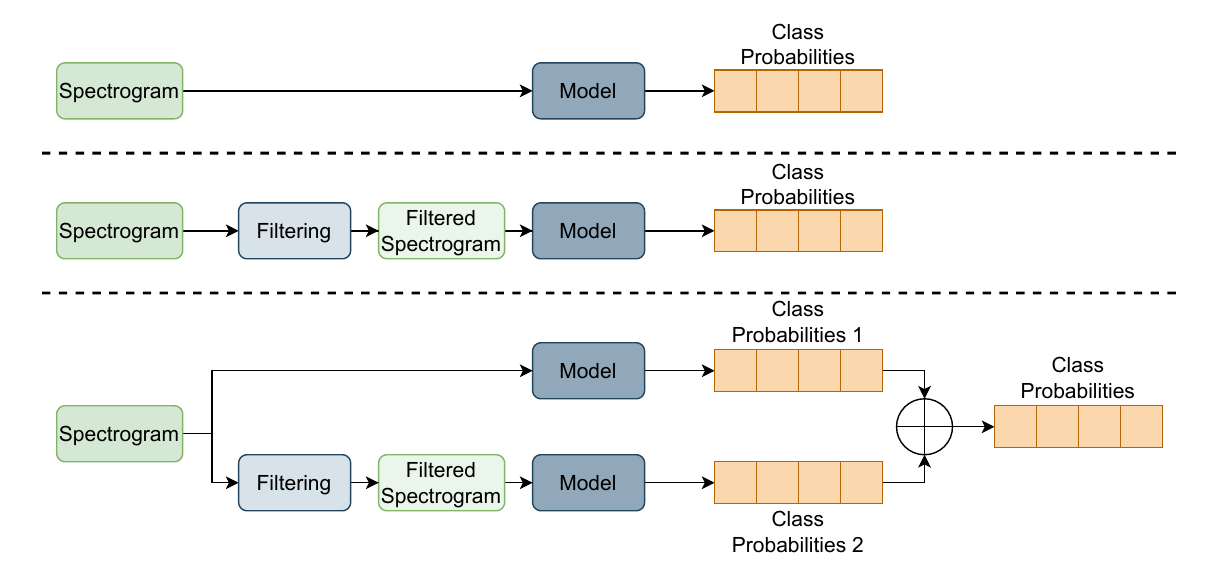}}
    \caption{Block diagram illustrating the use of the raw and filtered audios for spectrogram classification. The top row illustrates the classification pipeline using the unfiltered spectrogram, the middle row illustrates the use of filtering alone, and the bottom row illustrates the parallel classification using two instances of the model.}
    \label{fig:combining_models}
\end{figure}

The sound emitted by a ship is composed of machinery and propeller noises, which appear as constant lower frequencies components in the frequency spectra, and hydrodynamic noise, which is a broadband component that shifts between mid and high frequencies according to the ship's speed \parencite{carlton2018marine}. Considering ships as moving targets, the propagation of their radiated noise in the ocean is also affected by the LME. However, in this case, the effect contains valuable information about the depth of the source, which is intrinsically related to the ships' physical structure. To account for that, the LME is used as a complementary information in the classification pipeline, which considers the original audio together with the LME filtered version as inputs to the classification, as illustrated in \figureref{fig:combining_models}. 

A common strategy in the ship-type classification literature is to apply deep learning algorithms in spectrogram-like representations \parencite{domingos2022survey}. Knowing that the ship radiated noise is most present on the lower frequency spectra, the CQT is used to obtain the spectrogram. The time-filtering of the cepstral components was also performed in the CQT, aiming to evaluate the transferability of the strategy to other types of spectrogram and applications. As the CQT presents a logarithmic scale rather than a linear one, the cepstral representation obtained is not in the time domain as quefrencies, thus the components are denoted as cepstral coefficients. 

\subsubsection{Implementation Details}

The machine learning pipeline used for classification was adapted from \parencite{domingos2026improving}, which employs the GSE ResNeXt model due to its stability and fast convergence. To maintain a consistent training environment, the model was trained using the AdamW optimiser with a learning rate of $1\times10^{-3}$ and a weight decay of $1\times10^{-4}$ for regularisation. Training was conducted over 50 epochs with a mini-batch size of 64.

For deep learning classification tasks, dataset selection plays a crucial role, as models tend to replicate patterns and trends present in the training data. Additionally, large-scale datasets are typically required to ensure model convergence. Considering this need, the Ocean Networks Canada Society database\footnote{\url{https://oceannetworks.ca}}, a publicly available source of unlabelled underwater audio recordings, was selected. Due to the availability of Automatic Identification System (AIS) data for annotation, three time windows recorded between May 2016 and October 2018 at the Strait of Georgia are commonly used in the literature \parencite{irfan2021deepship, domingos2022investigation}. These recordings were acquired using an icListen smart hydrophone, with a bandwidth of 1Hz-12kHz and a sampling rate of 32,000Hz. The sensors were deployed at depths of 141m, 147m and 144m for each recording window, representing a shallow-water scenario with favourable conditions for LME occurrence. To expand the scope of the analysis and enable performance comparisons, the following two subsets of this data were used as reference:

\paragraph{DeepShip} Contains recordings annotated into four classes based on vessel type (tug, passenger ship, cargo ship, and oil tanker) \parencite{irfan2021deepship}. To avoid interference from overlapping ship signals, only recordings from a single vessel within a 2~km radius of the hydrophone were included in the database. If a second vessel entered this radius, the recording was truncated and the remainder discarded. The dataset comprises 613 recordings from 265 distinct ships, totalling approximately 47 hours of audio.

\paragraph{VTUAD v2} Using an automatic annotation pipeline \parencite{domingos2022investigation}, recordings from ten ship-type classes were collected: tug boat, passenger ship, cargo ship, oil tanker, pilot vessel, rescue, sailing, pleasure craft, fishing, and dredger. Unlike DeepShip, this dataset includes recordings from vessels within a 2 km radius of the hydrophone, under the condition that no other vessel is present within a 4 km radius. This constraint was applied to minimise interference from nearby vessels. The dataset contains 1,111 recordings from 479 different ships, with a total duration of approximately 116 hours.

The literature presents various approaches to the classification of underwater recordings. In this experiment, an analysis of different task configurations was conducted to provide a comprehensive evaluation of the filtering strategy’s performance.

\subsubsection{Dataset Segmentation Tasks}

In underwater acoustic classification, dataset preprocessing typically involves segmenting full-length audio recordings into smaller parts, which can then be used independently during training, testing, or validation stages. Three segmentation strategies are commonly applied \parencite{domingos2026improving, liu2021underwater}: \emph{Task 1} follows a segment-first approach, where the full audio is first segmented, the segments are randomised, and then divided into training and testing subsets; \emph{Task 2}, referred to as Sequential Audio Sampling (SAS), assigns consecutive portions of each audio file to either training or testing, followed by segmentation into smaller parts; \emph{Task 3}, referred to as Full-Length Audio Sampling (FAS), uses each full audio file as either a training or testing sample, which is then segmented into smaller parts. Due to the similarity between adjacent recordings, \emph{Task 1} is typically prone to overfitting \parencite{domingos2026improving}. Therefore, in this study, only \emph{Task 2} and \emph{Task 3} were considered, and are referred to as SAS and FAS, respectively.

\begin{figure}[ht]
    \centering
    \resizebox{\columnwidth}{!}{\includegraphics{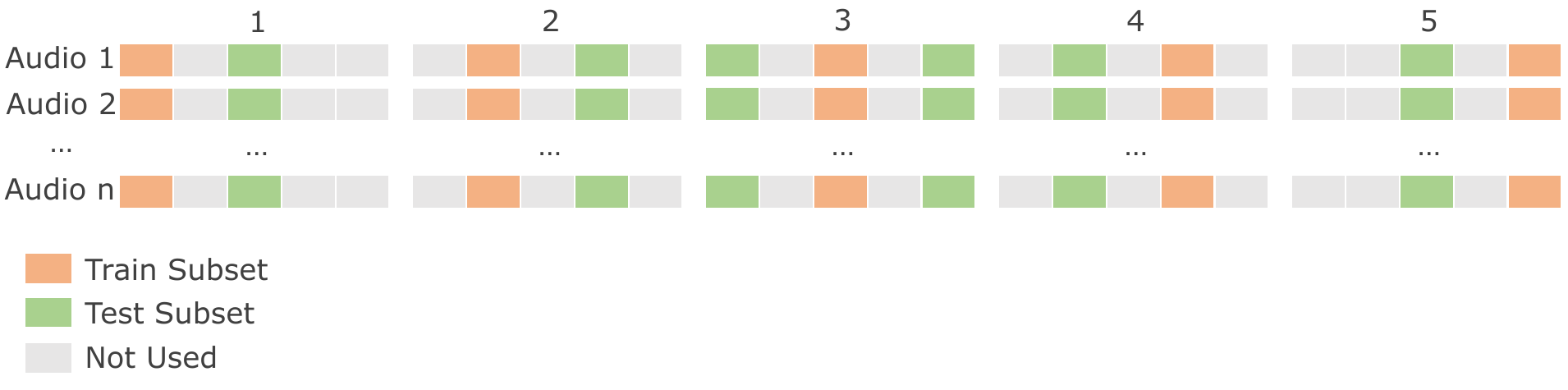}}
    \caption{Illustration of the five-fold division performed for SAS. The audio recordings, divided into five continuous segments, are shown five times across the rows to represent each fold. Segments used for training and testing are highlighted in different colours.}
    \label{fig:task2_folds}
\end{figure}

To ensure a comprehensive analysis of the results, a stratified 5-fold strategy was adopted for training and testing. In \emph{FAS}, each recording was assigned to a single fold, with training performed on four folds and testing on the remaining one. For \emph{SAS}, the fold division followed a strategy that avoids consecutive segments appearing in both training and testing subsets. This was achieved by selecting one major segment for training and another for testing, as illustrated in \figureref{fig:task2_folds}. The results for both \emph{SAS} and \emph{FAS} are reported in terms of the average and standard deviation across the 5-fold experiments.
 
\subsubsection{Classification Metrics}
In this research, classification performance is evaluated using the Matthews Correlation Coefficient (MCC) \parencite{matthews1975comparison}. The MCC is particularly suitable for handling class imbalances in classification tasks, as it assigns a high score to predictions that are accurate across all four base categories (true positives ($tp$), true negatives ($tn$), false positives ($fp$) and false negatives ($fn$)). The coefficient is defined as:

\begin{multline}
\text{MCC} = \\ \frac{tn\times tp - fn\times fp}{\sqrt{(tp + fp)\times(tp + fn)\times(tn + fp)\times(tn + fn) }}.
\end{multline}

Given that significant variations in classification performance are commonly observed across folds in underwater acoustic classification \parencite{domingos2026improving}, this research reports the MCC alongside the difference from the baseline (\(\Delta\)). This additional information enables direct comparison of the results with the baselines, while also accounting the impact of standard deviation in assessing statistical relevance, thereby reducing inconsistencies across folds.

\section{Results and Discussion}
\label{sec:results}

In this section, results from both experiments are presented. First, the attenuation of multi-path reflection components in synthetic data is analysed, followed by a discussion of the findings. The same structure is applied to the second experiment, which evaluates the effectiveness of the filtering approach in underwater acoustic signals within a classification pipeline.

\subsection{Multi-path Reflection Attenuation in Synthetic Data}

\begin{figure*}[ht]
    \centering
    \resizebox{\textwidth}{!}{\includegraphics{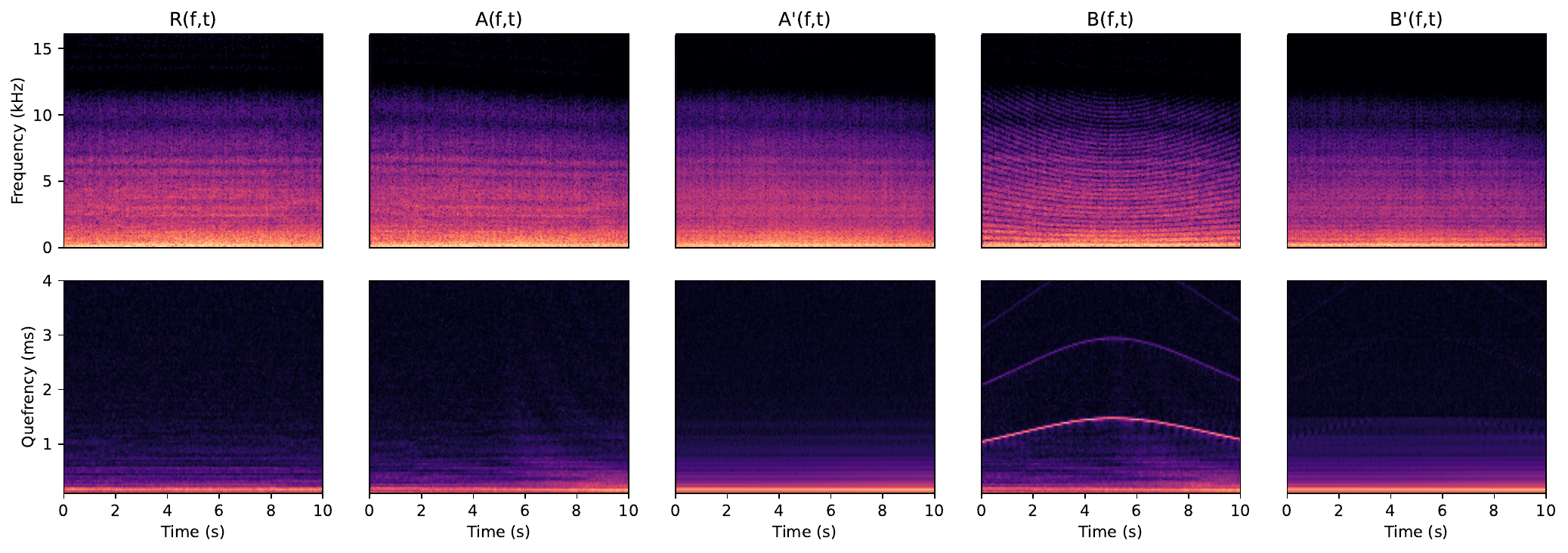}}   
    \caption[Comparison of the spectrograms and cepstrograms of the synthetic dataset components filtered using the time filtering of the cepstrum.]{Comparison of the spectrograms (upper row) and cepstrograms (bottom row) of the synthetic dataset components, filtered using temporal cepstral filtering. Reflection components are visible in the cepstrograms of $A(f,t)$ (at quefrencies below 1 ms) and $B(f,t)$ (at quefrencies between 1 and 2 ms), in contrast to $R(f,t)$. After filtering, these artefacts are attenuated in the cepstrograms of both $A'(f,t)$ and $B'(f,t)$, with the attenuation also observable in the corresponding recovered spectrograms.}
    \label{fig:synth_data_results}
\end{figure*}

Considering the dataset of acoustic data with simulated movement, the cepstral time-filtering was performed in the spectrograms using the adaptive band-stop filter. The filtering was applied in both the acoustic spectrographic data considering only the direct path component, here denoted as  $A(f,t)$, and the data considering the waves propagated trough direct and reflected paths, denoted as $B(f,t)$. Using the spectrogram of the acoustic data with amplitude attenuation as reference ($R(f,t)$), the similarity between the original signal and the reference was compared against the similarity between the filtered signals ($A'(f,t)$ and $B'(f,t)$) and the reference. \figureref{fig:synth_data_results} shows an example data containing the spectrograms and cepstrograms of the reference, delayed signals, and filtered acoustic signals.

\begin{figure}[ht]
    \centering
    \resizebox{\columnwidth}{!}{\includegraphics{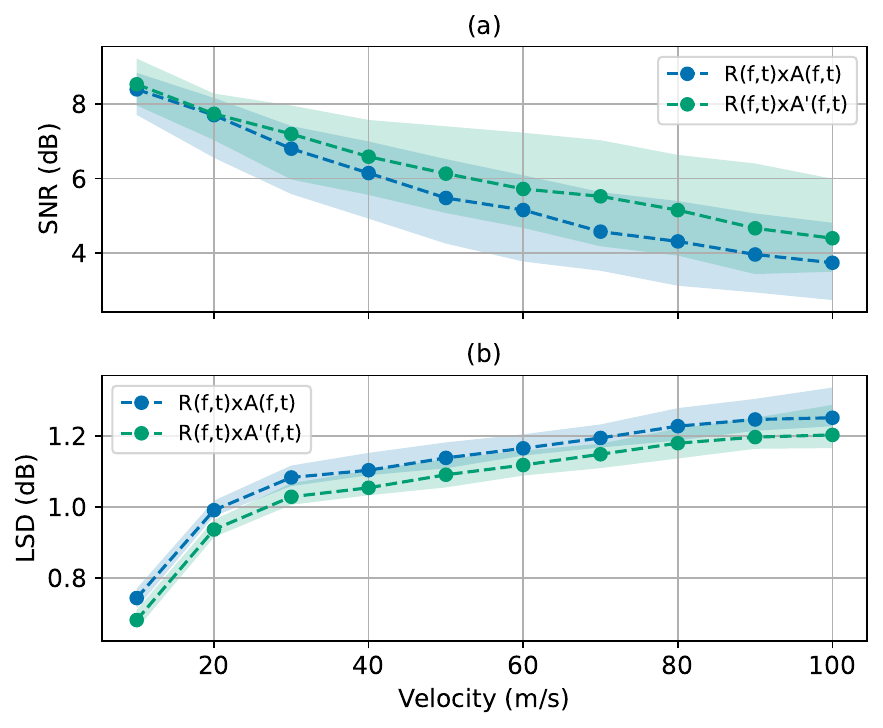}}   
    \caption{Similarity metrics obtained when comparing the reference signal $R(f,t)$ with a signal propagated along the direct path only ($R(f,t) \times A(f,t)$) and its filtered version ($R(f,t) \times A'(f,t)$), across different target velocities. Results are shown as the median value with shaded areas indicating the lower and upper quartiles. While cepstral filtering reduced spectral distortion, as measured by LSD, a reduction in SNR was noted in the absence of multi-path effects.}
    \label{fig:doppler_synt}
\end{figure}

The spectrograms containing the delay caused by the direct path of wave propagation ($A(f,t)$) exhibited an \emph{S-shape} curve localised at the simulated closest point of approach (see the frequency decaying around 5 seconds in \figureref{fig:synth_data_results}), posing as a challenge for the filtering algorithm due to the frequency shift caused by Doppler Effect. \figureref{fig:doppler_synt} shows the behaviour of the similarity and distortion metrics across all velocities for both $R(f,t) , A(f,t)$ and $R(f,t) , A'(f,t)$ pairs. As it is more sensitive to the absolute values, the SNR measurement showed no improvement in the ratio of signal of the filtered version in comparison with the unfiltered, as can be seen in graph (a). However, when considering the logarithmically compressed spectrogram in the LSD distortion metric, the filtering procedure demonstrated beneficial effects, as indicated by the lower distortion values obtained for the $R(f,t), A'(f,t)$ pair in the graphs shown in \figureref{fig:doppler_synt} (b). As it reflects the perceptual quality of the audio, the filtering process was successful in improving the quality of signals affected by the Doppler effect. As expected, the quality of both pairs decrease as the velocity increases, which reflects the intensity of the synthetic movement effects.

Discrepancies between the SNR measurements and LSD are related to the amplitude of the filtered signals. As SNR is sensitive to absolute values of magnitude, variations in the amplitude obtained after filtering have a substantial impact in the final metric \parencite{gray1980distortion}. As $A(f,t)$ only contains the directly propagated signal, as well as $R(f,t)$, the SNR between both will be similar, or sometimes higher, when compared with the filtered version, which can suffer from amplitude attenuation. For the distortion metric, which consider the ratio of the log components, the measurements are lower after the filtering process as the spectral shape is recovered. 

\begin{figure}[ht]
    \centering
    \resizebox{\columnwidth}{!}{\includegraphics{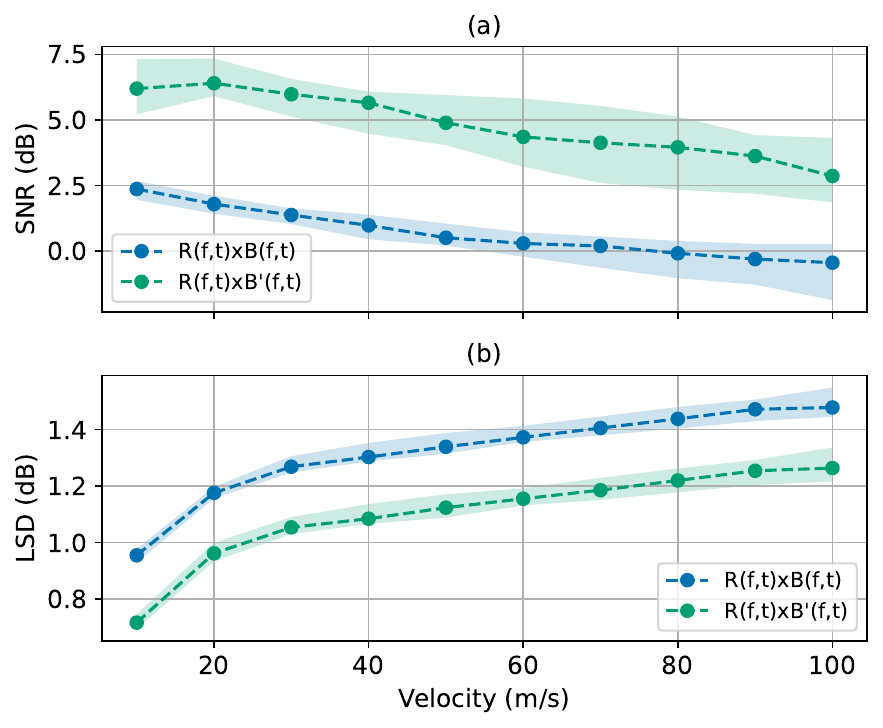}}   
    \caption{Similarity metrics obtained when comparing the reference signal $R(f,t)$ with a signal propagated along both direct and reflected paths ($R(f,t) \times B(f,t)$) and its filtered version ($R(f,t) \times B'(f,t)$), across different target velocities. Results are shown as the median value with shaded areas indicating the lower and upper quartiles. Cepstral filtering improved both LSD and SNR in the multi-path scenario.}
    \label{fig:lme_synt}
\end{figure}

The evaluation of the spectrogram containing both the waves propagated in the direct and reflected paths $B(f,t)$ and its filtered version $B'(f,t)$ exhibited positive results for the filtering strategy. As can be seen in \figureref{fig:lme_synt}, the comparison of the metrics curves between the two pairs of representations, $R(f,t) , B(f,t)$ and $R(f,t) , B'(f,t)$, showed a consistent quality improvement after the filtering process across all velocities. As the artefacts presented in $B(f,t)$ contemplate not only the Doppler effect but also the LME, the impact of the filtering strategy was more prominent, reflecting in both metrics, as can be seen in \figureref{fig:lme_synt} curves. Despite the increased distance between the curves in graph (b), the behaviour of the LSD was similar to that in the previous experiment, reflecting the improvements obtained after filtering. Also, graph (a) shows that the SNR of the filtered version was higher than the unfiltered, reflecting the capability of the filtering process of attenuating the echoes caused by the reflection components. 

The SNR values obtained in the second experiment had a greater increase after filtering when compared with the first experiment (see graphs (a) on \figureref{fig:doppler_synt} and \figureref{fig:lme_synt}). As reflections introduced constructive and destructive interference that altered the signal amplitude, their removal had a pronounced effect on SNR when comparison is made to a reference that did not contain reflected components. For LSD, the filtering process similarly leads to recovery of the spectral shape, reflecting an improvement in perceptual signal quality.

\begin{figure*}[ht]
    \centering
    \resizebox{\textwidth}{!}{\includegraphics{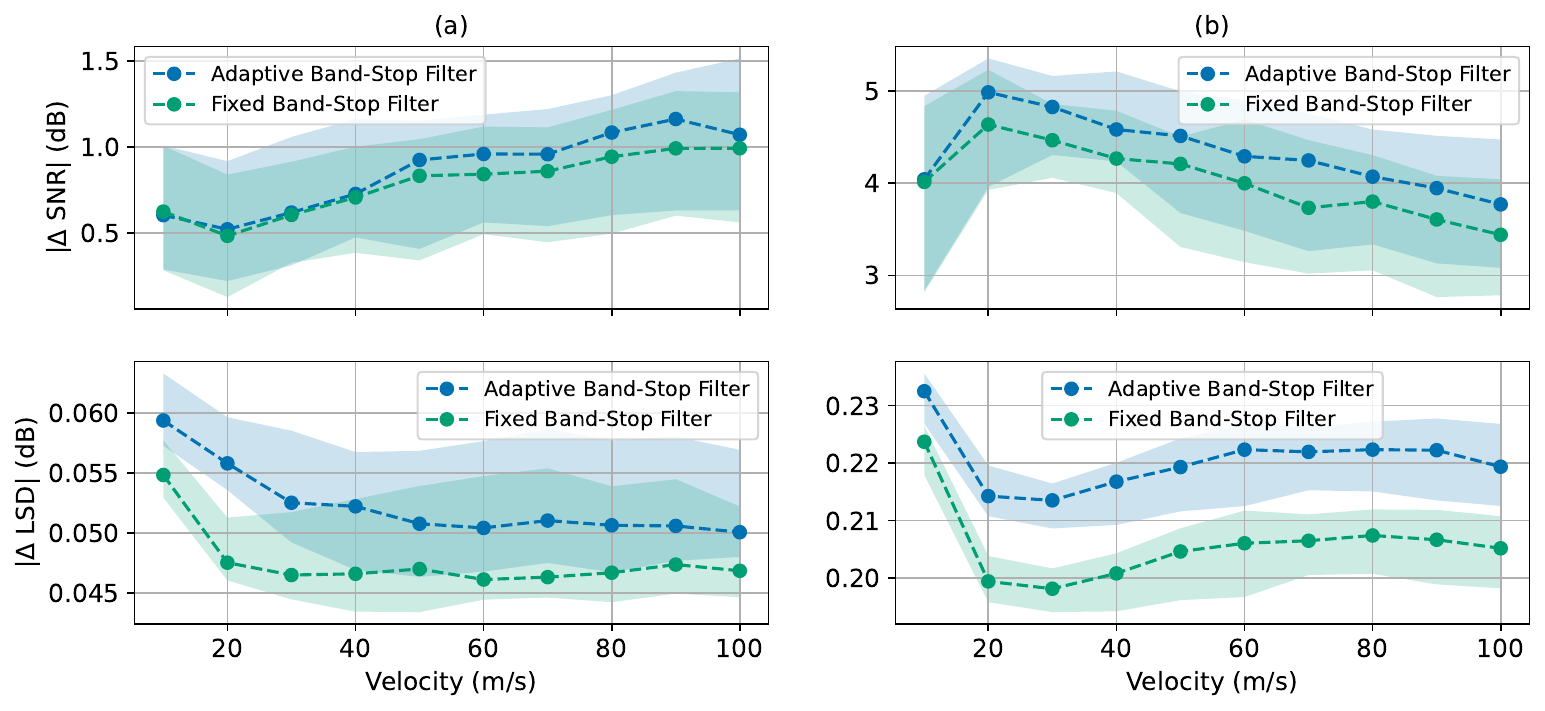}}   
    \caption{Comparison of metric differences ($\Delta$) between the reference and noisy signals ($R(f,t) \times A(f,t)$ and $R(f,t) \times B(f,t)$) and the corresponding filtered signals ($R(f,t) \times A'(f,t)$ and $R(f,t) \times B'(f,t)$), for both adaptive and fixed band‑stop filters. Column (a) shows results for the experiment involving propagation along the direct path only ($A(f,t)$), whereas column (b) corresponds to propagation along both direct and reflected paths ($B(f,t)$). Results are presented using the median value, with shaded regions indicating the lower and upper quartiles. Adaptive filtering leads to greater improvements in LSD in both experiments, particularly in the attenuation of multi‑path artefacts.}
    \label{fig:fix_vs_adp}
\end{figure*}

In addition to the previous experiments, a fixed band‑stop filter with an analogous frequency band was used as comparison to the adaptive band‑stop filter. To facilitate visualisation and comparison, only the absolute difference between the noisy and filtered metrics were considered. The results in \figureref{fig:fix_vs_adp} are summarised by the median across velocities, with shaded bands indicating the lower and upper quartiles. Both the direct‑path experiment (column (a)) and the multi‑path experiment including LME interactions (column (b)) are reported. 
In column (a), the median SNR shows modest gains for velocities above 50 m/s with the adaptive filter; however, these gains are not substantial once the quartiles are considered. A similar pattern is observed in column (b), with improvements across all velocities and reduced inter-quartile overlap, indicating a more pronounced improvement in SNR. For LSD, both experiments show improvements with adaptive filtering relative to fixed filtering. The effect is most pronounced in the multi‑path scenario (column (b)), with gains ranging from 4–8\% on the median value across velocities with no inter-quartile overlap.

Due to the dynamic nature of the target signature and the environment, adaptive filtering exhibited a better performance than fixed filtering. In multi‑path reflection scenarios, where signal variation is more pronounced, the benefit of adaptive filtering is most evident. The observed improvements in distortion metrics (e.g., LSD) indicate enhanced perceptual signal quality following filtering. The impact of these improvements on recognition and classification algorithms that operate directly on spectrogram‑based representations will be further evaluated.

\subsection{LME Filtering in Underwater Acoustic Classification}

\begin{figure}[ht]
    \centering
    \resizebox{\columnwidth}{!}{\includegraphics{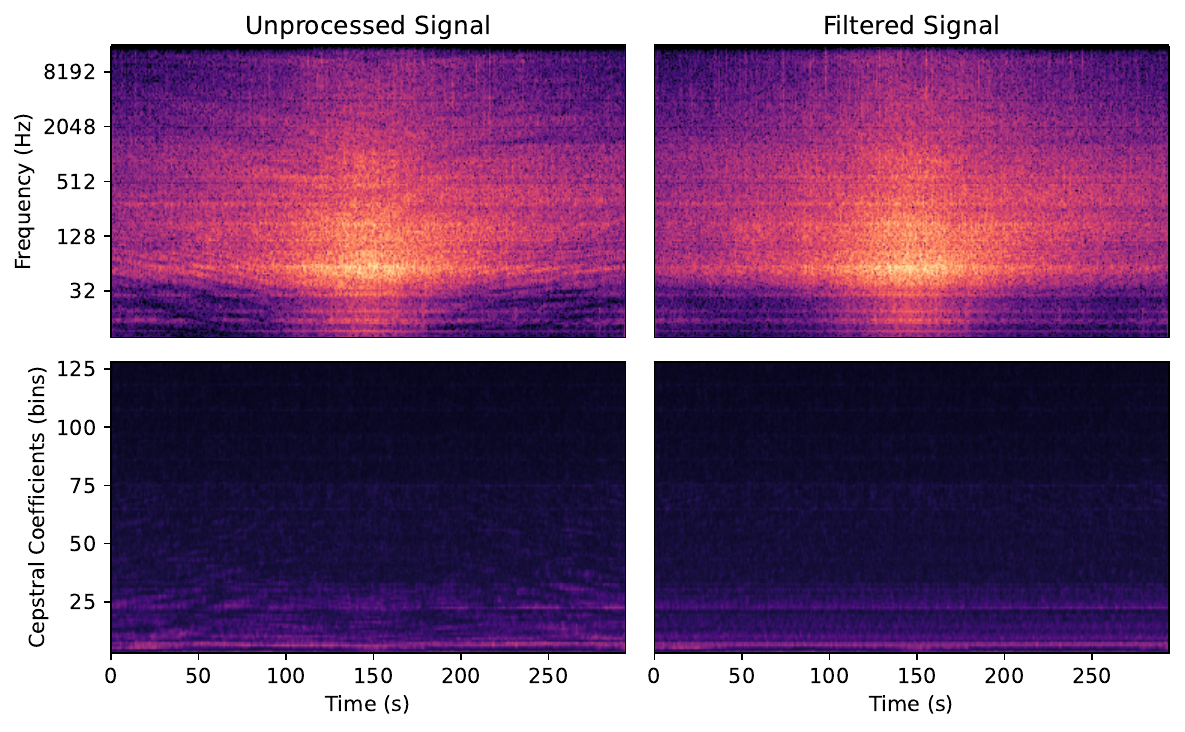}}   
    \caption{Comparison of the CQT spectrograms (top row) and cepstrograms (bottom row) of the unprocessed and filtered representations. In the unprocessed spectrogram signal, the u-shaped patterns are caused by the LME. After processing, only the straight horizontal lines are maintained, which represent the harmonic components of the ship-radiated noise.}
    \label{fig:cepstrum_filt}
\end{figure}

The filtering method was applied to the CQT cepstral coefficients of the underwater acoustic data to attenuate the LME present in CQT spectrograms. \figureref{fig:cepstrum_filt} shows the qualitative attenuation of LME and movement artefacts after the filtering process. In the CQT spectrogram of the unprocessed signal, the LME effect is visible at lower frequencies, resulting from reflections of the sound wave on the seafloor. This pattern is also evident in the cepstral coefficients, appearing as diagonal lines in the lower coefficients (around the bin 25 in \figureref{fig:cepstrum_filt}). The behaviour of the CQT cepstral coefficients is analogous to that observed in cepstrograms, confirming that reflection patterns are identifiable in this representation.

After filtering, both the unprocessed and filtered signals were used as inputs to a classification pipeline. The results, presented in \tableref{table:baseline_tasks_classification}, show the performance of the classification algorithm across five scenarios: A classification pipeline using the unprocessed spectrogram alone; a pipeline which uses the spectrogram after the time-filtering of the cepstrum for both adaptive and fixed band-stop filtering; and the combination of the unprocessed and the filtered spectrograms in a parallel classification for both filtering strategies, as illustrated in \figureref{fig:combining_models}.

\begin{table*}[tb]
\caption{Classification performance for SAS and FAS on both datasets over the 5-fold experiment, reported with standard deviation. The difference (\(\Delta\)) from the unprocessed baseline is included for reference.}
\label{table:baseline_tasks_classification}
\centering
	\begin{tabular}{llcccc}
	\hline
        \multirow{2}[1]{*}{\textbf{Dataset}} & \multirow{2}[1]{*}{\textbf{Input}} & \multicolumn{2}{c}{\textbf{SAS}} & \multicolumn{2}{c}{\textbf{FAS}}\\
	  & & \textbf{MCC (\%)} & \textbf{$\Delta$} & \textbf{MCC (\%)} & \textbf{$\Delta$}\\
	\hline
	\multirow{5}[1]{*}{\textbf{DeepShip}} & Unprocessed & 78.09 ± 3.96 & -- & 66.45 ± 5.02 & --\\
        & Fixed Filtered alone & 77.18 ± 2.03 & -0.91 ± 2.17 & 66.08 ± 4.94 & -0.37 ± 3.03\\
        & Adaptive Filtered alone & 77.65 ± 2.46 & - 0.44 ± 3.14 & 65.56 ± 5.08 & - 0.89 ± 0.95\\
        & Unprocessed + Fixed & 79.62 ± 2.86 & + 1.53 ± 1.25 & 68.16 ± 4.74 & + 1.71 ± 1.04\\
	    & Unprocessed + Adaptive & 80.37 ± 2.70 & + 2.28 ± 1.30 & 67.79 ± 5.19 & + 1.35 ± 0.54\\
	\hline
	\multirow{5}[1]{*}{\textbf{VTUAD v2}} & Unprocessed & 79.71 ±1.51 & -- & 51.51 ±0.68 & --\\
        & Fixed Filtering & 79.80 ± 1.28 & 0.09 ± 1.09 & 50.03 ± 0.99 & -1.47 ± 0.96\\
        & Adaptive Filtered alone & 79.26 ± 1.49 & - 0.45 ± 0.69 & 50.59 ± 0.62 & - 0.92 ± 0.96\\
        & Unprocessed + Fixed & 82.55 ± 1.32 & + 2.84 ± 0.74 & 52.23 ± 0.83 & + 0.72 ± 0.34\\
	      & Unprocessed + Adaptive & 82.33 ± 1.71 & + 2.62 ± 0.54 & 52.36 ± 0.30 & + 0.85 ± 0.50\\
	\hline
	\end{tabular}
\end{table*}

The results indicate that using the filtered representation alone does not improve classification performance for SAS and FAS on either dataset. Consistent with this, the average differences (\(\Delta\)) are smaller than their respective standard deviations, indicating that the filtered representation alone does not yield consistent improvements. On the other hand, the parallel filtering pipeline, which combines filtered and unprocessed representations, yields noticeable improvements in classification performance across both tasks and datasets. SAS showed the most significant gains, with increases of 3.46\% and 3.29\% for DeepShip and VTUAD v2, respectively, when using the adaptive filtering. FAS also benefited from the parallel approach, with improvements of 2.03\% and 1.65\% for DeepShip and VTUAD v2, respectively. The differences between fixed and adaptive filtering are generally subtle, often within one standard deviation. An exception occurs for SAS under the parallel pipeline, where adaptive filtering surpasses fixed filtering by a small margin.

As expected, applying filtering in isolation was not beneficial. The filtering strategy proved effective in attenuating reflected components; however, it also suppressed information encoded in those components, such as the sound source depth and range \parencite{ferguson2019improved}. This information can correlate with target characteristics (e.g., ship length, tonnage) and thus with vessel type, making them potentially discriminative \parencite{sys2008search}. Therefore, while the filter successfully mitigates the influence of multi-path reflections, it may also remove discriminative features that are valuable for classification.

Since SAS utilises different segments of the same audio for training and testing, the attenuation of time-varying components proved beneficial for classification. In contrast, FAS benefited more from the complementary information provided by the LME. Because the acoustic source in a ship is related to its physical dimensions, such as the draught and the propeller position, the LME pattern indirectly provides information about the ship's size and draft\parencite{sys2008search ,johnson2025characterization}. Retaining this information in the unprocessed signal contributed positively to classification. However, as the engine and machinery sounds were hidden in the reflection artefacts \parencite{audoly2017measurement}, the filtering strategy contributed to the identification of harmonic patterns present in the spectrograms. In practice, the parallel classification pipeline provides both representations separately to the model, offering complementary perspectives that enhance performance.

\section{Conclusion}
\label{sec:final_remarks}
This research introduced an innovative temporal filtering approach for cepstrogram representations to attenuate multi-path reflections and motion-induced artefacts. The proposed method incorporates an adaptive band-stop filter that dynamically adjusts its bandwidth based on the relative intensity of quefrency components. This framework successfully separated reflected components from the source signal, improving the spectrogram representation of acoustic data.

The performance of the filtering method was evaluated through two experiments: (i) attenuation of multi-path reflections and motion artefacts in aircraft noise with simulated movement; and, (ii) application to a practical ship-type classification task employing underwater acoustic signals. In the simulated scenario, the filtering strategy effectively recovered the harmonic structure of the target signal across a range of velocities, mitigating both motion artefacts and multi-path effects. In environments affected by the LME, metrics like SNR and LSD improved across all evaluated velocities. Also, the improvement over a fixed band filter was between 4-8\% across velocities. In the classification task, the proposed filtering led to improvements in the MCC of 3.46\% and 3.29\% for the DeepShip and VTUAD V2 datasets, respectively.

Despite the strong results in attenuating motion artefacts, the SNR was practically maintained in cases where no reflected components were present, suggesting that the filtering process may alter signal amplitude. This highlights the need for future improvements in amplitude preservation. Nevertheless, the harmonic structure of the signal was enhanced, as evidenced by a reduction in the LSD ranging from 15\% to 25\% across different velocities. Additionally, the adaptive filtering strategy can represent a computational overhead in comparison with the fixed band filter, needing evaluation on time-restricted applications.

These findings demonstrate the effectiveness of the proposed pipeline in attenuating acoustic motion artefacts and separating multi-path reflections from the original source signal. The filtering strategy has the potential to enhance a range of applications, including acoustic target classification, time-delay estimation, target recognition, and underwater communication in multi-path environments.
Future research should focus on improving amplitude conservation after filtering and exploring the application of this method in towed array systems and multi-sensor configurations.

\section*{Acknowledgments}
This work was supported by the National Industry PhD Program, Australia, in a partnership between Flinders University and PrioriAnalytica Pty Ltd.

\bibliographystyle{IEEEtran}
\bibliography{references}

\newpage
\vspace{11pt}

\vfill
\end{document}